\documentclass[aps,pre,twocolumn,superscriptaddress,floatfix]{revtex4}
\usepackage{graphicx}
\usepackage[dvips,unicode,colorlinks,linkcolor=blue,citecolor=blue,urlcolor=blue]{hyperref}
\begin{document}
\title{Accelerating oscillatory fronts in a nonlinear sonic vacuum with strong non-local effects}

 \author{O. V. Gendelman}
 \email{ovgend@tx.technion.ac.il}
 \affiliation{
 Faculty of Mechanical Engineering, Technion -- Israel Institute of Technology,
 Haifa 32000, Israel}

 \author{V. Zolotarevskiy}
 \affiliation{
 Faculty of Mechanical Engineering, Technion -- Israel Institute of Technology, Haifa 32000, Israel}

 \author{A. V. Savin}
 \affiliation{Semenov Institute of Chemical Physics, Russian Academy of Sciences, Moscow 119991, Russia}

 \author{L. A. Bergman}
 \affiliation{Department of Aerospace Engineering, University of Illinois in Urbana-Champaign, USA}

 \author{A. F. Vakakis}
 \affiliation{Department of Mechanical Science and Engineering,
              University of Illinois in Urbana-Champaign, USA}

\date{\today}

\begin{abstract}
In this Letter we describe a novel class of dynamical excitations -- accelerating oscillatory fronts
in a new genre of nonlinear sonic vacua with strongly non-local effects. Indeed, it is surprising that such
models naturally arise in dynamics of common and popular lattices. In this study, we address a chain of particles oscillating in the plane
and coupled by linear springs, with fixed ends. When one end of this system is harmonically excited in the transverse
direction, one observes accelerated propagation of the excitation front, accompanied by an almost
monochromatic oscillatory tail. The front propagation obeys the scaling law $l \sim t^{4/3}$.
The frequency of the oscillatory tail remains constant, and the wavelength scales as $\lambda \sim t^{1/3}$.
These scaling laws result from the nonlocal effects; we derive them analytically (including the scaling coefficients) from a continuum approximation.
Moreover, a certain threshold excitation amplitude is required in order to initiate the front propagation.
The initiation threshold is rationalized on the basis of a simplified discrete model. This model is further reduced to a new completely integrable nonlinear system. The Letter introduces a new and yet unexplored class of nonlinear sonic vacua and explores the effects of strong non-locality on the initiation and propagation of oscillating fronts in these media. Given their simplicity, nonlinear sonic vacua of the type considered herein should be common in periodic lattices.
\end{abstract}


\maketitle

Nonlocal nonlinearities naturally appear in classical models describing the nonlinear motion of
beams with restrained ends \cite{LL,NAY}. For example, the transverse dynamics of a beam with fixed ends, bending in a direction of one of the
main axes of its cross-section, is described by the following partial integro-differential equation:
\begin{equation}
\label{eq1}
%\rho \ddot{y}+EIy^{IV}=ES{y}''(T+\frac{1}{2L}\int\limits_0^L {{y}'^{2}dx}).
\rho y_{tt}+EIy_{xxxx}=ESy_{xx}\left( T+\frac{1}{2L}\int\limits_0^L y_x^2dx\right).
\end{equation}
Here $y(x,t)$ is the transverse displacement of the beam point; $x$ is a coordinate along its axis;
$\rho$ and $E$ are the mass density and Young's modulus, respectively, of the beam; $L$ is the length of the
undeformed beam; $S$ and $I$ denote area and moment of inertia of the beam cross-section and
$T$ is an applied axial tensile force in $x$ direction. The nonlocal term in the right-hand side of Eq.(\ref{eq1})
appears due to a stretching of the midplane caused by transverse displacement of the beam with
immobile ends. If the bending term is negligible, the axial tensile force is absent, and both
ends are clamped at distance $L$; then Eq.(\ref{eq1}) reduces to the following strongly
nonlinear wave equation:
\begin{equation}
\label{eq2}
%\rho \ddot{y}-\frac{ES}{2L}{y}''\int\limits_0^L {{y}'^{2}dx} =0.
\rho y_{tt}-\frac{ES}{2L}y_{xx}\int\limits_0^L y_x^{2}dx =0.
\end{equation}
This equation describes the transverse oscillations of an elastic string with fixed ends and without
pre-tension. Only recently, a similar approximation has been developed for a discrete counterpart
of such a string in the form of a linear chain of particles moving in the plane, with next-neighbor interactions through linear springs, and with fixed boundaries \cite{MV}.
As shown in \cite{MV} and Supplemental Material \cite {SM}, if the transverse displacements of the chain of particles are not too
large compared to the length of the free springs, the transverse motion of the chain of particles
decouples from the longitudinal motion, and is approximately described by the following set
of ordinary differential equations:
\begin{equation}
\label{eq3}
%\ddot{y}_{n} +\frac{2y_{n} -y_{n-1} -y_{n+1} }{2(N-1)}\sum\limits_{j=1}^{N-1} {(y_{j+1} -y_{j} )^{2}}=0.
\ddot{y}_{n} -\frac{y_{n+1}-2y_{n} +y_{n-1}}{2(N-1)}\sum\limits_{j=1}^{N-1} {(y_{j+1} -y_{j} )^{2}}=0.
\end{equation}
In this system $y_n$ is the transverse displacement of the $n$-th particle, $N$ -- the overall number
of the particles, $y_{1}=y_{N}=0$. The mass of each particle and the spring constant are set to unity
without loss of generality. It is easy to recognize that Eq. (\ref{eq2}) approximates Eq.(\ref{eq3})
in the long-wave limit.

Systems (\ref{eq2}) and (\ref{eq3}) exemplify the important concept of {\it sonic vacuum} --
linearization of both these systems yields zero sound velocity. Thus, these systems can be
classified as {\it nonlinearizable} -- one never can neglect the nonlinear terms. Similar properties
of the sonic vacuum are well-known and widely studied in systems of granular particles without
external pre-compression; the nonlinearity appears there due to Hertzian contact \cite{G1,G2,G3,G4}.
Also, nonlinearizable systems of different structure were widely explored as possible nonlinear
energy sinks \cite{S1,S2,S3,S4}. The essential nonlinearity of these sinks allows them to resonate
with primary oscillatory systems over a broad frequency range \cite{S2} and efficiently absorb energy.
Such systems are investigated as possible engineering solutions for vibration mitigation and
energy harvesting in a wide range of possible applications \cite{S2,A1,A2}.

However, systems (\ref{eq2}), (\ref{eq3}), as compared to the granular media without pre-compression or
the nonlinear energy sinks, possess an additional important property -- the nonlocality. If one is
interested only in modal oscillations of these systems, the problem becomes relatively easy if one notes that the
integral term in Eq. (\ref{eq2}) and the corresponding sum term in Eq. (\ref{eq3}) depend only on
time. Therefore, the spatial modal shapes of the strongly nonlinear sonic vacua  (\ref{eq2}) and (\ref{eq3}) will be the same as for simple linear string or chain with
fixed ends, and, respectively, will correspond to sinusoidal standing waves. The nonlinearity will reveal
itself only in the time domain -- the oscillations will be anharmonic, and their frequency will be
proportional to the amplitude \cite{MV}. The fact that the integral term in Eqs. (\ref{eq1}), (\ref{eq2})
does not affect the spatial modal shapes, is well-known and widely used in approximate modal
analysis of oscillating continuous systems \cite{Go1}.

Beyond  modal oscillations, the dynamics of sonic vacua similar to (\ref{eq2}), (\ref{eq3}) is almost unexplored
due to the strong nonlinearity. Below we demonstrate that the nonlocal nonlinearity brings
about some unexpected and unusual dynamic phenomena even in very simple settings. To illustrate that,
the dynamics of a spring-and-mass chain with $N$ particles in a plane is simulated. This system is detailed in the Supplemental Material (system (1), \cite{SM}), and in the limit of low energy its transverse oscillations are approximately described by Eq. (\ref{eq3}). The non-dimensional
Hamiltonian of the lattice is expressed as
\begin{equation}
H=\sum_{n=2}^{N-1}\frac12(\dot{x}_n^2+\dot{y}_n^2)+\sum_{n=1}^{N-1}\frac12(r_n-1)^2,
\label{f1}
\end{equation}
where $x_n$ and $y_n$ denote the axial and transverse coordinates of the of the $n$-th particle, $r_n=[(x_{n+1}-x_n)^2+(y_{n+1}-y_n)^2]^{1/2}$
is the distance between particles $n$ and $n+1$. The right end of the chain is fixed, whereas the left end of
the chain does not move in the $x$ direction, and harmonically oscillates in $y$ direction.
The equations of motion are expressed as
\begin{eqnarray}
\label{f2}
x_1(t)\equiv 0,~~ y_1(t)=A\sin(\omega t), \nonumber\\
\ddot{x}_n=-\partial H/\partial x_n, ~~ \ddot{y}_n=-\partial H/\partial y_n, ~~ n=2,...,N-1, \\
x_N(t)\equiv N-1,~~y_N(t)\equiv 0. \nonumber
\end{eqnarray}
and are numerically integrated with zero initial conditions by the velocity
Verlet method \cite{VM} with the following parameters: $N\in[125,10000]$, excitation amplitudes $A\in[0.1,7]$
and frequencies $\omega\in[0.01,0.5]$. Small viscous friction is imposed on a few rightmost
particles to improve the simulation accuracy and avoid numerical discrepancies.
%-------------------------- FIG 1 ------------------------------------------
\begin{figure}[tb]
\includegraphics[angle=0, width=0.8\linewidth]{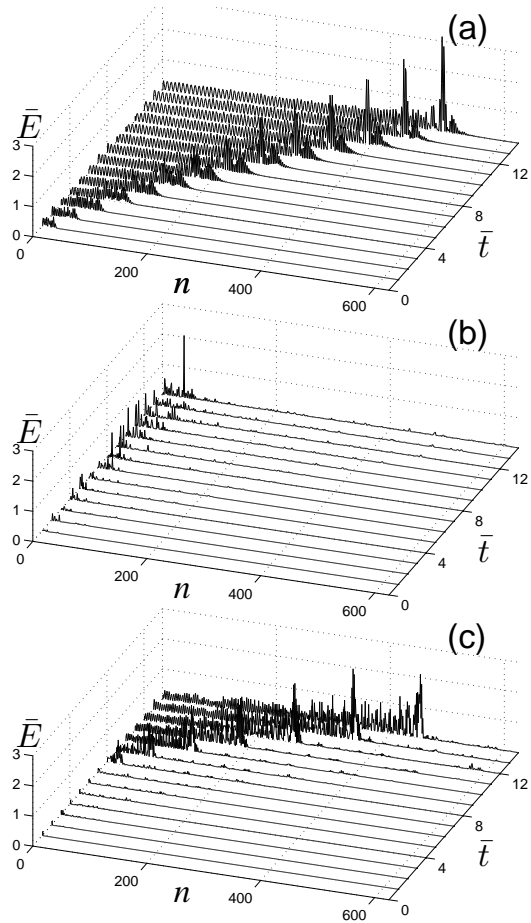}
\caption{
Evolution of local kinetic energy ${\bar E}=\dot y_n^2/2\Delta E$ at different segments of time
$\bar t=t/\Delta t$ in the chain with $N=625$ and (a) $A=0.5$, $\omega=0.05$ ($\Delta E=10^{-3}$, $\Delta t=492$); (b) $A=1$, $\omega=0.3$ ($\Delta E=1$, $\Delta t=164$);
(c) $A=0.15$, $\omega=0.05$ ($\Delta E=10^{-4}$, $\Delta t=2048$).
}
\label{fig01}
\end{figure}
%-------------------------- FIG 1 ------------------------------------------

\begin{figure}[tb]
\includegraphics[angle=0, width=0.8\linewidth]{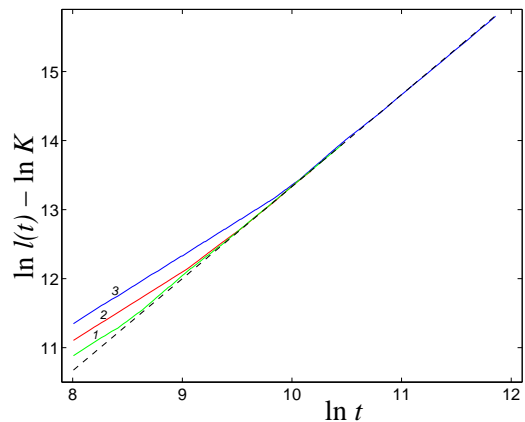}
\caption{(Color online)
Dependence of the position of the leading edge of the front $l(t)$ on time $t$
for three different sets of parameters:
$N=2500, A=0.5, \omega=0.03$; $N=5000, A=1.5, \omega=0.08$ and
$N=1000, A=0.6, \omega=0.03$ (curves 1, 2 and 3 respectively).
Dashed line corresponds to the slope $t^{4/3}$.
}
\label{fig02}
\end{figure}

Typical results of the simulation are presented in Figure \ref{fig01}, where we depict the "transverse component"
of local kinetic energy ${E_k}=\dot y^2_n/2$ versus $n$ at different time instances.
One can observe the propagation of the excitation front, accompanied by an apparently monochromatic
oscillatory tail. Moreover, it is clear from Figure \ref{fig01} (a), that this front accelerates
in the course of propagation. It is interesting to note that the stationary fronts with oscillatory tails, but with constant velocity, are
well-known in models of phase transitions in solid state and similar problems \cite{P04,M90,SGM14,SR15}.
Moreover, it is obvious that in a linear wave equation with similar boundary excitation one would
observe the monochromatic oscillatory front propagating with the sound velocity. It follows that the observed
acceleration of the front should be attributed to the nonlocality of the nonlinear term in (\ref{eq3}).
In order to explain this finding analytically, we consider a simplified model of the oscillatory
region in the chain and suppose a monochromatic wave in the oscillatory tail after the front in continuum approximation.
Only transverse oscillations are taken into account (numerical justification of this assumption is presented in the Supplemental Material \cite{SM}). Then, the field of displacements in the
oscillatory zone is described as follows:
\begin{equation}
\label{eq4}
y(x,t)=\left\{ {\begin{array}{l}
A\sin (\omega t-kx),~~0\le x\le l(t), \\
0,~~l(t)<x\le L. \\
\end{array}} \right.
\end{equation}
Here $l(t)$ is the instantaneous coordinate of the front, and $k$ is the wavenumber. It is also assumed
that the front propagation is slow enough compared to the frequency of transverse oscillations of the particles; i.e., $dl(t)/dt\ll\omega l(t)$.
To establish complete correspondence between the continuum approximation (\ref{eq2}) and the discrete model
(\ref{eq3})--(\ref{f2}), one should set $\rho=E=S=1, L=N-1$. Then, by substituting (\ref{eq4})
to (\ref{eq2}) and balancing principal terms, the following equation is obtained:
\begin{equation}
\label{eq5}
A\omega^{2}\sin (\omega t-kx)\approx
\frac{k^{4}A^{3}l(t)\sin (\omega t-kx)}{4(N-1)}.
\end{equation}
An additional condition can be obtained from the assumed stationary character of the front propagation.
To this end, the phase velocity of the oscillatory tail should be equal to the front velocity \cite{SGM14} so that
$V_{ph}=\omega/k=dl(t)/dt$. Combining this expression with Eq.(\ref{eq5}), one obtains
an explicit expression for the position of the accelerating front:
\begin{eqnarray}
\label{eq6}
k^{4}l(t)=\frac{4\omega^{2}(N-1)}{A^{2}}\Rightarrow
\frac{1}{l}(\frac{dl}{dt})^4=\frac{A^{2}\omega^{2}}{4(N-1)}\Rightarrow \\ \nonumber
\Rightarrow
l(t)=K t^{4/3},~~ K=\left( {\frac{81A^{2}\omega^{2}}{1024(N-1)}} \right)^{1/3}
\end{eqnarray}
Thus, the front indeed accelerates with velocity $V \sim t^{1/3}$. Prediction of Eq. (\ref{eq6})
is completely supported by the numerical simulations, as is demonstrated in Figure \ref{fig02} --
for three different sets of parameters, with the curves depicting the front position versus time (
shifted by $\ln K$) collapsed into a straight line with slope $4/3$. So, the considerations presented
above predict not only the correct scaling law for the front position, but also the scaling coefficient,
which depends on specific set of parameters.
%-------------------------- FIG 2 ------------------------------------------
%-------------------------- FIG 2 ------------------------------------------
%-------------------------- FIG 3 ------------------------------------------
\begin{figure}[tb]
\includegraphics[angle=0, width=0.8\linewidth]{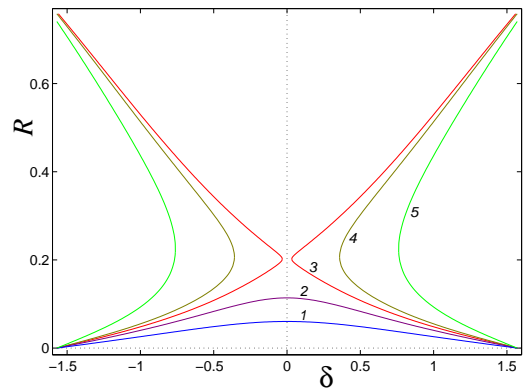}
\caption{(Color online)
Phase trajectories corresponding to Equation (\ref{eq10})for zero initial conditions and
$a=0.3,0.34,0.3575,0.37,0.42$ (curves 1, 2, 3, 4 and 5 respectively).
}
\label{fig03}
\end{figure}
%-------------------------- FIG 3 ------------------------------------------

The simulated system is discrete rather than continuous, and that is why not every set of
parameters leads to formation of the accelerating front.
This point is illustrated in Fig. \ref{fig01} (b):
If the excitation amplitude is too small, the oscillations remain localized at the left end of the chain.
This phenomenon cannot be explained in terms of the continuum model (\ref{eq2}).
To describe the front initiation above a certain excitation threshold, we resort to the discrete model of the
chain. We will adopt a simplified approach and establish the minimal
amplitude of oscillations of particle $n=1$ that allows efficient
excitation of particle $n=2$ and, thus, substantial excitation of the chain and initiation of the wave front.
Accordingly, we analyze (\ref{eq3}) with
$y_{1} =- A\sin \omega t,y_{n} = 0,n\geq 3$ and zero initial
conditions for $y_{2} (t)$. This system is rescaled with
$\omega t=\tau$, $y_{2} =\alpha u(\tau )$, $A=\alpha a$, $\alpha =\omega \sqrt {2(N-1)}$.
Then one arrives at the following equation for variable $u(\tau )$:
\begin{equation}
u_{\tau \tau } +(2u+a\sin \tau )[u^{2}+(u+a\sin \tau )^{2}]=0.
\label{eq7}
\end{equation}
The primary frequency of the oscillatory front is expected to be close to the normalized value of unity. Therefore,
the complex variable $\varphi (\tau )\exp (i\tau )=u_{\tau } (\tau )+iu(\tau )$ is introduced \cite{MG}.
Supposing that variable $\varphi (\tau )$ varies slowly; balancing
principal terms in Eq. (\ref{eq7}), we arrive at the following slow-flow equation:
\begin{eqnarray}
\varphi_{\tau } +\frac12 i\varphi -\frac18 i[2(2\varphi +a)(\left|
\varphi \right|^{2}+\left| {\varphi +a} \right|^{2})+ \\ \nonumber
+(2\varphi^{\ast}+a)(\varphi^{2}+(\varphi +a)^{2})]=0.
\label{eq8}
\end{eqnarray}
Though far from obvious, this slow-flow equation is completely integrable. The integral of motion is
expressed as:
\begin{equation}
C=\left| \varphi \right|^{2}-\frac{1}{8}[2(\left| \varphi \right|^{2}+\left|
{\varphi +a} \right|^{2})^{2}+\left| {\varphi^{2}+(\varphi +a)^{2}}
\right|^{2}].
\label{eq9}
\end{equation}
To see this, it is sufficient to note     that Eq. (\ref{eq8}) is equivalent to
$\varphi_{\tau } =-\frac{i}{2}\frac{\partial C}{\partial \varphi^{\ast }}$.
Since $C$ is real, one immediately obtains
$\varphi_{\tau }^{\ast } =\frac{i}{2}\frac{\partial C}{\partial \varphi }$
and, consequently,
$\frac{dC}{d\tau }=\frac{\partial C}{\partial \varphi }\varphi_{\tau }
+\frac{\partial C}{\partial \varphi^{\ast }}\varphi_{\tau }^{\ast } =0$.
Then we split the slow variable into polar components
$\varphi =R\exp (i\delta )$. Initial condition $\varphi (0)=0$ corresponds to $C=-3a^{4}/8$.
Therefore, the averaged phase
trajectories of the particle $n=2$ for different values of the external
excitation are expressed by the following family of implicit equations:
\begin{equation}
R^{2}(1-\frac{3}{2}R^{2}-a^{2})-3Ra(R^{2}+
\frac12a^2)\cos \delta-2(Ra\cos\delta)^{2}=0.
\label{eq10}
\end{equation}
%-------------------------- FIG 4 ------------------------------------------
\begin{figure}[tb]
\includegraphics[angle=0, width=0.75\linewidth]{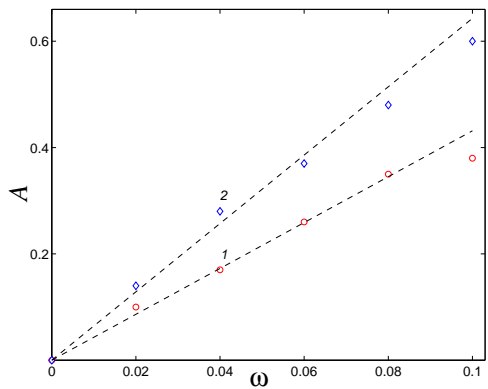}
\caption{(Color online)
Correlation between the frequency $\omega$ and the amplitude $A$
for the wave front for chain lengths $N=125$  and
$N=250$ (lines 1 and 2). The black dashed lines represent linear fitting of the markers 1 and 2.
The region above a corresponding dashed line corresponds to a propagating front regime,
and no front is initiated for parameters below the dashed line.
The linear fitting relations are given by $A=0.274\omega\sqrt{2(N-1)}$
for $N=125$ and $A=0.287\omega\sqrt{2(N-1)}$ for $N=250$.
}
\label{fig04}
\end{figure}
%-------------------------- FIG 4 ------------------------------------------

The family of solutions of Equation (\ref{eq10}) for various values of $a$ is
presented in Fig.~\ref{fig03}. One can see that for small values of $a$ the
phase trajectory stays in the region of small values of $R$. This regime corresponds to
localization of forced oscillations near the excited end of the chain, similar to the regime
demonstrated in Fig.~\ref{fig01} (b). There
exists an excitation threshold above which the phase trajectory is
attracted to a region of relatively large $R$. So, the energy of
oscillations is intensively irradiated into the chain, and it seems natural to
associate this regime with the formation of the oscillatory front. The threshold
excitation corresponds to the phase trajectory, which passes through the saddle point in
Fig.~\ref{fig03} (curve 3). This yields the following evaluation for
this threshold:
\begin{equation}
a_{cr}\approx 0.3574
\label{eq11}
\end{equation}
So, the boundary for formation of the oscillatory front is described by the line $A=a_{cr} \omega \sqrt {2(N-1)} $. This
prediction is verified in Figure \ref{fig04}. Approximate
linear dependence of A on $\omega$ is observed, but the
 coefficient of these lines turns out to be somewhat overestimated.

This discrepancy is obviously related to the number of simplifying
assumptions adopted in our analysis. Besides, it is apparent from Fig.~\ref{fig01} (b),
that localized oscillations at the left end of the chain exhibit chaotic dynamics.
Then, it is possible to expect that the appropriate conditions for the front
initiation may be formed also if the excitation is far below the threshold --
just due to fluctuations. An example of such behavior --  the front formation is observed after a
certain time delay -- is presented in Fig.~\ref{fig01} (c).

To conclude, we revealed a new type of excitations in a lattice representing a nonlinear sonic vacuum with
strong nonlocal dynamical interactions (despite only next-neighbor physical coupling). These excitations are accelerating
fronts with oscillatory tails. The fronts accelerate according to the scaling
law $l \sim t^{4/3}$ due to nonlocal dynamical interactions. The tails have constant frequency, but their
wavelength is  not constant -- it scales with time as
$\lambda =2\pi/k=2\pi l_t/\omega \sim t^{1/3}$.
Such fronts reveal themselves in most well-known and popular models, such
as the suspended string without pre-tension and the chain of linear springs and
masses with fixed ends. Simple analytic considerations allow derivation of
all main parameters of the front, including the scaling characteristics and
the excitation threshold. Due to the fixed boundary conditions, such fronts
can exist only as transient regimes. At the same time, due to extreme
simplicity and popularity of the involved models, one can expect to see
such accelerating fronts in many physical settings.

The authors are very grateful to the Israel Science Foundation (grant 838/13)
for financial support.


\begin{references}
\bibitem{LL}
L.D.Landau and E.M.Lifshitz, {\it Theory of Elasticity} (Pergamon Press, 1970)

\bibitem{NAY}
A.H.Nayfeh, {\it Nonlinear Interactions} (Wiley, 2000).

\bibitem{MV}
L.I.Manevitch and A.F.Vakakis,
%Nonlinear Oscillatory Acoustic Vacuum,
SIAM Journal of Applied Mathematics, {\bf 74}, 1742 (2014).

\bibitem{SM}
See Supplemental Material at [URL will be inserted by publisher]

\bibitem{G1}
V.F. Nesterenko, {\it Dynamics of Heterogeneous Materials} (Springer Verlag, New York, 2001).

\bibitem{G2}
C. Daraio, V.F. Nesterenko, E.B. Herbold, and S. Jin,
%Tunability of solitary wave properties in one-dimensional strongly nonlinear phononic crystals,
Phys. Rev. E {\bf 73}, 026610 (2006).

\bibitem{G3}
S. Sen, J. Hong, J. Bang, E. Avalos, and R. Doney,
%Solitary waves in the granular chain,
Phys. Rep., {\bf 462}, 21 (2008).

\bibitem{G4}
Y. Starosvetsky, M.A. Hasan, A.F. Vakakis, and L.I. Manevitch,
%Strongly nonlinear beat phenomena and energy exchanges in weakly coupled granular chains on elastic foundations,
SIAM Journal of  Applied Mathematics, {\bf 72}, 337 (2012).

\bibitem{S1}
O.V. Gendelman,
%Transition of Energy to a Nonlinear Localized Mode in a Highly Asymmetric System of Two Oscillators,
Nonlinear Dynamics {\bf 25}, 237 (2001).

\bibitem{S2}
A.F. Vakakis, O.V. Gendelman, G. Kerschen, L.A. Bergman, D.M. McFarland, and Y.S.
Lee, {\it Nonlinear Targeted Energy Transfer in Mechanical and Structural Systems} (Springer
Verlag, Berlin, 2008).

\bibitem{S3}
L.I. Manevitch, E. Gourdon, and C.H. Lamarque,
%Towards the design of an optimal energetic sink in a strongly inhomogeneous two-degree-of freedom system,
Journal of Applied Mechanics, {\bf 74}, 1078 (2007).

\bibitem{S4}

G. Sigalov, O. V. Gendelman, M. A. Al-Shudeifat, L. I. Manevitch, A. F. Vakakis and L. A. Bergman,
 % Alternation of regular and chaotic dynamics in a simple two-degree-of-freedom system with nonlinear inertial coupling,
 Chaos {\bf 22}, 03318 (2012).

\bibitem{A1}
R. Bellet, B. Cochelin, P. Herzog, and P.-O. Mattei,
%Experimental study of targeted energy transfer from an acoustic system to a nonlinear membrane absorber,
Journal of Sound and Vibration, {\bf 329} 2768 (2010).

\bibitem{A2}
Z. Nili Ahmadabadi and S.E.Khadem,
%Nonlinear vibration control of a cantilever beam by a nonlinear energy sink
Mechanism and Machine Theory, {\bf 50}, 134 (2012).

\bibitem{Go1}
E. Hollander and O.Gottlieb,
%elf-excited chaotic dynamics of a nonlinear thermo-visco-elastic system that is subject to laser irradiation
Applied Physics Letters, {\bf 101}, 133507 (2012).

\bibitem{VM}
L. Verlet
%Computer "Experiments" on Classical Fluids. I. Thermodynamical Properties of Lennard-Jones Molecules
Phys. Rev. {\bf 159}, 98 (1967).

%1
\bibitem{P04}
D. Panja, Phys. Rep. {\bf 393}, 87 (2004).
\
%2
\bibitem{M90}
A. S. Mikhailov, {\it Foundations of Synergetics 1. Distributed Active
Systems} (Springer-Verlag, Berlin, 1990).
%3
\bibitem{SGM14}
V.V. Smirnov, O.V. Gendelman, and  L.I. Manevitch,  Phys. Rev. E
{\bf 89} 050901(R) (2014).
%19
\bibitem{SR15}
M. Schaeffer and M. Ruzzene, Int. J. Solids Struct. {\bf 56-57}, 78 (2015).
\bibitem{MG}
L.I. Manevitch and O.V. Gendelman, {\it Tractable Models of Solid Mechanics} (Springer Verlag,
Heidelberg, 2011).
\end{references}
\end{document}